# An Improved CVM Entropy Functional for BCC Alloys

Vikas Jindal*, Shrikant Lele

*Centre of Advanced Study, Department of Metallurgical Engineering,*
*Institute of Technology, Banaras Hindu University,*
*Varanasi 221 005, India*

*corresponding author E-mail address: vjindal.met@itbhu.ac.in (Vikas Jindal)

Abstract

We explore the possibility of modifying the multiplicity of the basic cluster in the entropy functional used in the cluster variation method so that truncation errors owing to finite size of the basic cluster may be corrected. The numerical value of this multiplicity is found by requiring the modified CVM entropy functional (M-CVM) in the tetrahedron approximation for the body-centered cubic structure to yield the exact critical temperatures for ordering and phase separation. We demonstrate that very accurate values of the order parameter, correlation functions and the Gibbs function can be obtained for ordered and disordered phases at arbitrary composition and temperature with M-CVM approach by comparing the results with those for Monte Carlo simulations.

## 1. Introduction

A quantitatively reliable description of the thermodynamics of disordered and ordered phases present in alloys and their equilibria is essential for the development of newer alloys. The Monte Carlo (MC) method [1] and the cluster variation method (CVM) [2, 3] are the methods most often applied to bulk alloy systems for this purpose.

Although the MC technique is more accurate, it involves time consuming calculations and is thus restricted to coarse grids of chemical potential and temperatures. Further, MC simulations do not directly give the values of important thermodynamic functions such as entropy and the Gibbs function, since these quantities cannot be expressed in terms of ensemble averages. Instead, these are obtained by integration of thermodynamic relations from a known starting point [1].

On the other hand, the CVM proposed by Kikuchi [2] is simple and reproduces most of the topological features of phase diagrams. However, lower approximations of CVM are less accurate than Monte-Carlo (MC) simulations. Ferreira et al [4] have shown that correlation functions for face-centered cubic (fcc) structures determined using the tetrahedron approximation of CVM underestimate the exact values found using MC simulation. This leads to an overestimation of entropy.

One may also adopt a hybrid approach in which correlation functions are obtained from MC simulations and then used in the CVM expression for entropy and the Gibbs function to achieve greater accuracy [4, 5, 6]. This approach retains the simplicity of the CVM and the accuracy and flexibility of MC. This method reduces to some extent the computational burden associated with the MC method since thermodynamic integration is replaced by CVM for obtaining entropy. However, the computational effort still remains high. This hybrid method has been applied to fcc ordering [4] and disordered phase separating [7] systems and to body-centered cubic (bcc) systems [6]. With a view to increase the accuracy of the hybrid MC-CVM method, Ferreira et al. [4] have suggested a new functional for the entropy that reproduces the exact MC entropy very closely. Higher accuracy can also be achieved by using a larger cluster approximation for writing the CVM entropy expression [7] again resulting in high computational effort. A slightly different approach has been taken by Oates et al. [8], who have treated the coefficient of the term corresponding to the basic cluster in the entropy functional as an adjustable system dependent parameter in addition to the effective cluster interaction energies to be determined by fitting to experimental phase diagram data in a version of CVM designated cluster site approximation (CSA).

The object of this study is to present a modified CVM entropy functional which does not use any MC results but closely reproduces the MC correlation functions and entropy. Kikuchi and Brush [9] proposed that in formulating the CVM entropy hierarchy the dimensionality can be reduced by one, that is a hierarchy of (D-1)-dimensionally extended clusters can be used in improving the entropy of a D-dimensional system, with D being 2 or 3. The convergence of the hierarchy to the rigorous limit was proved by Schlijper [10] for the 2-D case and by Kikuchi [11] for both the 2-D and 3-D cases based on a conjecture. However, there is an exponential increase in complexity and computational effort with increasing size of the basic cluster. Limiting the basic cluster size thus leads to a truncation of the entropy functional. In such a situation, it is possible to modify the term corresponding to the basic cluster in the entropy functional to reproduce some desirable feature of the exact behavior. We shall apply this approach to bcc alloys in the irregular tetrahedron approximation of CVM. The modified entropy functional has been validated by comparison with MC simulations because they provide accurate values of the correlation functions for known effective pair interaction energies. On the other hand there are no experimental methods for determining correlation functions except for pair correlation functions which are related to the Cowley-Warren short range order parameters [3]. However, even the latter cannot be used for validation since the effective pair interaction energies are not known a priori for any real system. Section 2 gives the procedure for the modification of multiplicity of basic cluster and details of MC simulations performed for the present work. Results for modified multiplicity for the maximal cluster are given in section 3. Further, various thermodynamic functions obtained from modified Gibbs function are compared with MC simulations in this section. Finally, the present work is evaluated against existing alloy models in section 4.

## 2. Methodology

The CVM entropy functional counts the maximal and all sub-clusters as many times as its multiplicity in the structure. To correct for the effect of truncation, we consider a modification of the multiplicity of the basic cluster leaving the multiplicities of all subclusters unchanged. As a result the Kikuchi-Barker (K-B) coefficients of the latter are also modified. The modification of the multiplicity of the basic cluster is tailored to yield the exact value of the consolute temperature in phase separating systems as well as the critical order-disorder transition temperature in ordering systems. To try out this idea, we consider the case of a binary bcc alloy in the irregular tetrahedron approximation of CVM.

The clusters are numbered in the following sequence for the disordered structure: 1. I-n Pair, 2. II-n Pair, 3. Triangle, 4. Tetrahedron and 5. Point. For the ordered B2 structure, we distinguish sites on two sublattices $\alpha$ and $\beta$. This leads to the following clusters: 1. I-n $\alpha\beta$ pair, 2. II-n $\alpha\alpha$ and $\beta\beta$ pairs, 3. Triangles of the types $\alpha\alpha\beta$ and $\alpha\beta\beta$, 4. Tetrahedra of the type $\alpha\alpha\beta\beta$ and 5. Points of the types $\alpha$ and $\beta$. The K-B coefficients, $\gamma$, of the subclusters are calculated following the usual procedure [3] in terms of the modified multiplicity $m_4$ of the tetrahedron cluster as follows. The superscripts refer to the sublattices associated with the cluster sites for the ordered B2 structure.

$$\gamma_1 = \gamma_1^{\alpha\beta} = -5 + m_4 \tag{1}$$

$$\gamma_2 = \gamma_2^{\alpha\alpha} = \gamma_2^{\beta\beta} = -3 + 2m_4/3 \tag{2}$$

$$\gamma_3 = \gamma_3^{\alpha\alpha\beta} = \gamma_3^{\alpha\beta\beta} = 1 - m_4/3 \tag{3}$$

$$\gamma_4 = \gamma_4^{\alpha\alpha\beta\beta} = 1 \quad (4)$$

$$\gamma_5 = \gamma_5^{\alpha} = \gamma_5^{\beta} = 23 - 4m_4 \quad (5)$$

### *2.1. Consolute Point of Phase Separating System*

Consider a phase separating system with repulsive I-n pair interactions. At the consolute point of such a system, the second derivative of the Gibbs function with respect to composition, i.e., $G_{xx} = (\partial^2 G/\partial x_B^2)_T$ vanishes [12] and thus

$$\left.\left(\frac{\partial^2 G}{\partial x_B{}^2}\right)_T\right|_{T=T_c} = 0 \quad (6)$$

For choosing the modified value of multiplicity, the $G_{xx}$ has been evaluated at the exact consolute temperature ($T_c$ = 6.3552 kT/J) and composition ($x_c$ = 0.5) [13] for different values of the multiplicity around the initial value of 6. For small changes in the multiplicity, the plot of $G_{xx}$ vs $m_4$ is nearly linear and it is possible to determine the value of $m_4$ for which $G_{xx}$ vanishes.

### *2.2. Critical Point of Ordering System*

Now consider an ordering system with attractive I-n pair interactions. At the critical point of such a system an order–disorder transition of second order occurs in bcc structures and the second derivative of the Gibbs function with respect to the long range order parameter $\eta$, i.e., $G_{\eta\eta} = (\partial^2 G/\partial \eta^2)_T$ vanishes [12] and thus

$$\left.\left(\frac{\partial^2 G}{\partial \eta^2}\right)_T\right|_{T=T_c} = 0 \quad (7)$$

A procedure similar to that described above for phase separating systems has been adopted for determining the modified multiplicity of tetrahedra.

### *2.3. Monte Carlo Simulations*

In order to show the effect of this change in the multiplicity on the correlation functions and to compare them with exact values, we have performed MC simulations and obtained values of all the correlation functions and also of the order parameter in case of the ordered B2 structure. These correlation functions were then utilized in the hybrid CE-CVM (using the modified CVM entropy functional) to obtain the configurational entropy. All the simulations were performed on a $16^3$ MC cell with 1000 MC steps per site (MCSS) discarded, and averages taken over the subsequent 4000 MCSS separated by 2 performed MCSS to ensure uncorrelated equilibrium configurations. Twenty such simulations were performed and the standard deviation of the average values of the correlation functions was found and reported as error.

## 3. Results

### *3.1. Modified multiplicity of tetrahedra*

The procedure explained in section 2.1 has been applied at the exact consolute temperature and composition to determine modified values of multiplicity for phase separating systems. A plot of the $G_{xx}$ versus multiplicity of the tetrahedron cluster is shown in fig. 1. It may be observed from fig. 1 that equation (6) is not satisfied at the initial value (namely 6) of multiplicity of tetrahedra. The $G_{xx}$ became exactly zero for the multiplicity of tetrahedra being 5.70017 (correct up to six significant digits).

A similar plot of the second derivative of $G_{\eta\eta}$ versus multiplicity of the tetrahedron cluster is also shown in fig. 1 for ordering systems. Again, the same value of 5.70017 for multiplicity satisfies equation (7). An identical change in the multiplicity of the tetrahedron cluster for both cases could be expected since there is no frustration in the bcc structure.

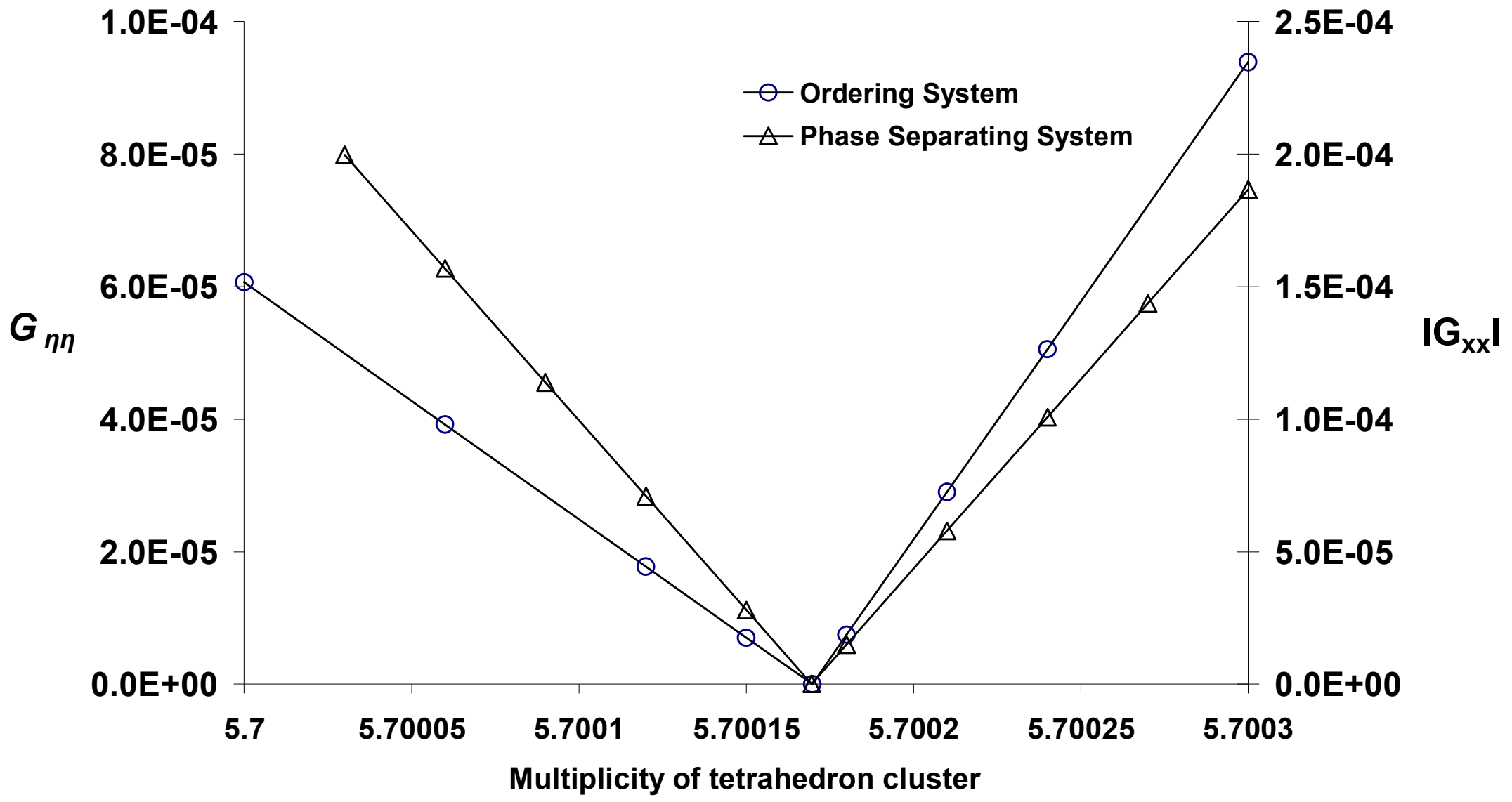

Fig. 1: Variation of $G_{xx}$ and $G_{\eta\eta}$ for A2 and B2 phases respectively with multiplicity of the tetrahedron cluster

### *3.2. Comparison with MC simulations*

In this section we will present results obtained by using modified CVM (M-CVM) entropy functional that is by using a multiplicity of 5.70017 for tetrahedra and compare them with those obtained using standard CVM and MC simulation for ordering as well as phase separating systems.

#### *3.2.1. Effect of change in temperature*

Fig. 2 illustrates the variation of the values of the order parameter and the correlation functions as a function of temperature for an equiatomic alloy in an ordering system using all three methods. For the disordered A2 phase, it was observed by Ferreira et al. [4] that CVM correlation functions are more random as compared to MC simulations. A similar trend is observed even in this investigation. However an opposite tendency is observed for ordered phase where correlation functions calculated using CVM are less random compared to those obtained from MC simulations. These differences become negligible far away from the transition temperature and may be attributed to mean field approximation of the Hamiltonian in case of CVM. However, M-CVM entropy functional not only corrects the transition temperature but also significantly improves agreement with the exact MC correlation functions. The largest deviations are observed at the transition temperature. These are recorded in table 1 along with the MC error (σ). For ordering as well as phase separating bcc systems, the values of correlation functions (including the Bragg-Williams long range order (LRO) parameter for the ordering case) found from M-CVM entropy functional are within ±3σ error bound of the values found from MC simulations. This represents a considerable improvement in comparison with CVM calculations.

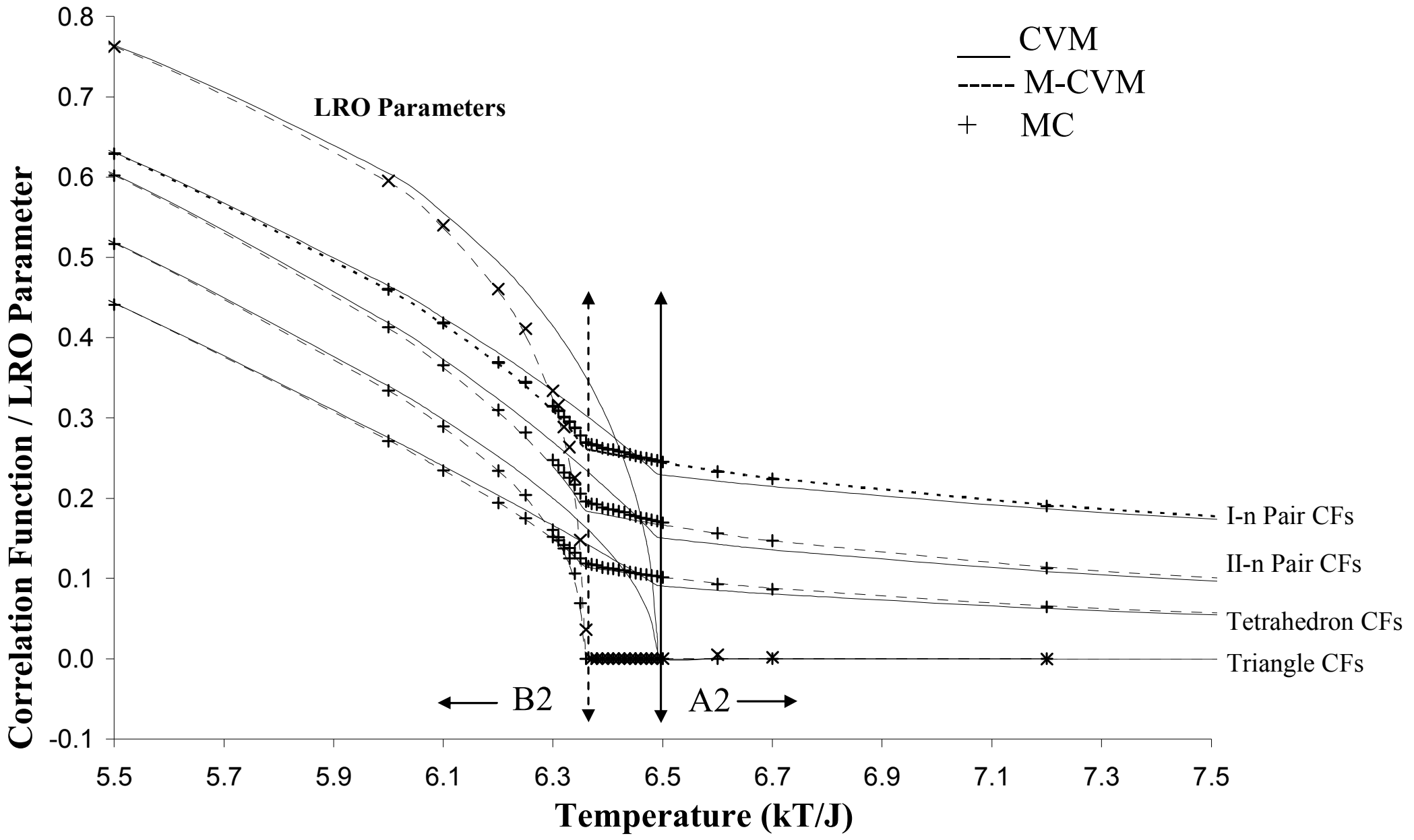

Fig. 2: Variation of correlation functions and long range order parameter with temperature for B2 and A2 phases of equiatomic composition

Since MC simulation does not provide the value of entropy directly, the correlation functions obtained from the MC simulations have been used in the M-CVM entropy functional and entropy values have been calculated. These are shown in fig. 3. Here also very good agreement has been achieved between results obtained from MC simulation and M-CVM entropy functional.

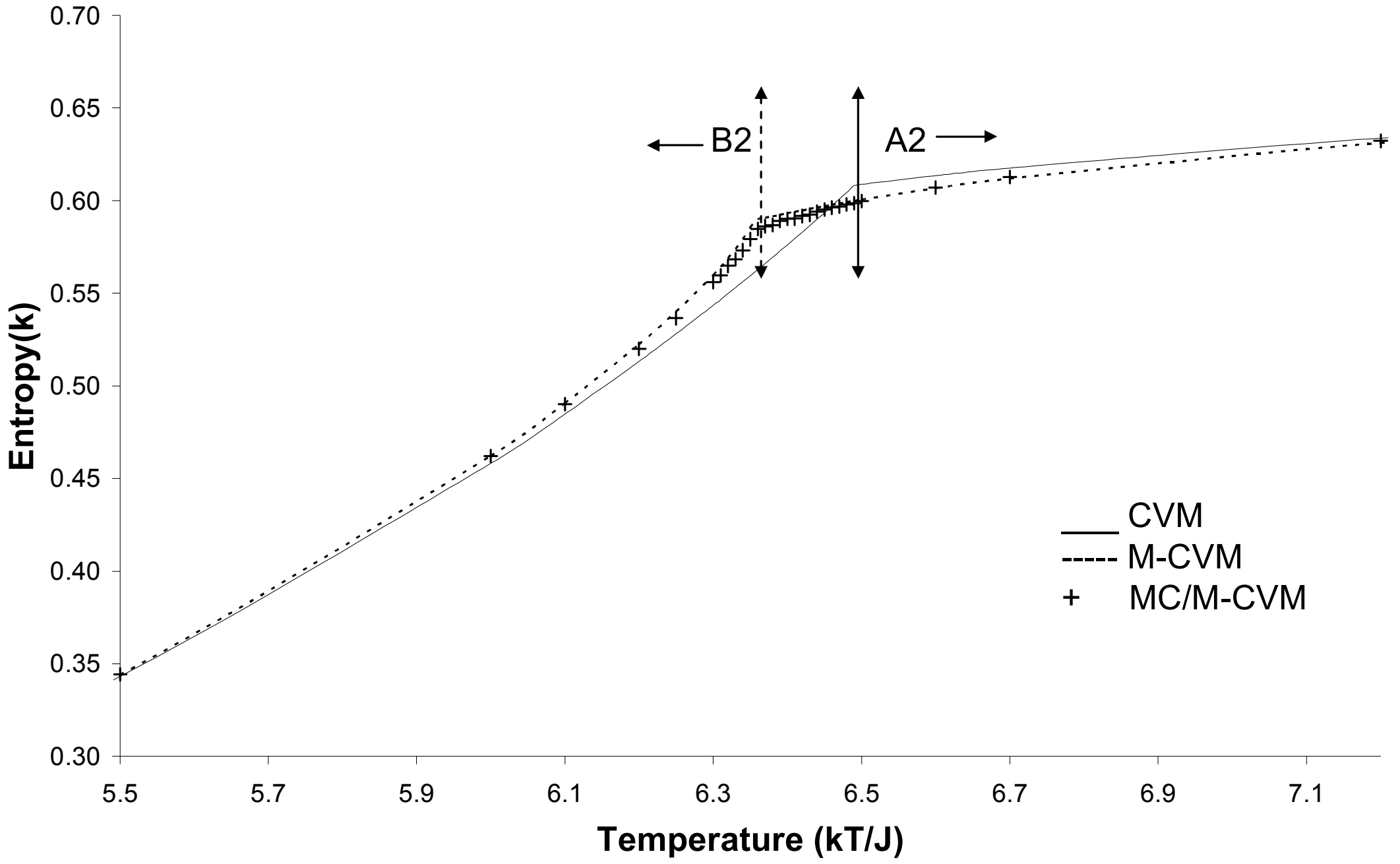

Fig. 3: Variation of entropy with temperature for B2 and A2 phases of equiatomic composition

Table 1. Comparison of values of LRO parameter and correlation functions (CF) at exact $x_c$, $T_c$ for ordering and phase separating systems

| Structural variable | CVM | M-CVM | MC | Error |
|---|---|---|---|---|
| Ordering system | | | | |
| LRO parameter | 0.3565 | 0.0 | 0.0000 | 0.0002 |
| I-n pair CF | -0.3055 | -0.2618 | -0.2570 | 0.0027 |
| II-n pair CF (αα) | 0.2377 | 0.1849 | 0.1816 | 0.0032 |
| II-n pair CF (ββ) | 0.2377 | 0.1849 | 0.1816 | 0.0031 |
| Triangle CF (ααβ) | 0.1664 | 0.0 | 0.0001 | 0.0003 |
| Triangle CF (αββ) | 0.1664 | 0.0 | 0.0001 | 0.0003 |
| Tetrahedron CF | 0.1451 | 0.1147 | 0.1101 | 0.0019 |
| Phase separating system | | | | |
| I-n pair CF | 0.2413 | 0.2618 | 0.2575 | 0.0019 |
| II-n pair CF | 0.1628 | 0.1849 | 0.1822 | 0.0021 |
| Triangle CF | 0.0 | 0.0 | 0.0000 | 0.0003 |
| Tetrahedron CF | 0.0994 | 0.1147 | 0.1105 | 0.0014 |

### *3.2.2. Effect of change in composition*

To understand the effect of change in the composition on the values of correlation functions and entropy calculated using M-CVM entropy functional we have chosen a temperature of 6.2 kT/J for ordering as well as phase separating systems. For the ordering case, the system shows presence of an ordered phase for equiatomic composition separated by the A2/B2 phase boundary from the disordered phase for compositions near the pure components. Variation of the order parameter and all the correlation functions with composition is shown in fig. 4. The A2/B2 phase boundary is apparent near $x_B$ = 0.417, 0.439 and 0.432 for the CVM, M-CVM and MC models respectively. It may be noted that the M-CVM phase boundary composition is closer to the exact MC composition. Again, when compositions are far away from the transition composition, all three models show similar values of the correlation functions. The largest deviations are observed at the transition composition. These are recorded in table 2 along with the MC error. For a phase separating system at $T$ = 6.2 kT/J, spinodal compositions are found to be 0.341, 0.338 and 0.339 for CVM, M-CVM and MC models respectively. Since these compositions are very close, all three models are compared at $x_B$ = 0.340 in table 2. It is evident from table 2 that, in general, values of correlation functions found from M-CVM calculations lie within error bounds of those found from MC simulations although the agreement here is not as good as that at the exact consolute or critical point. The largest deviation occurs for the I-n correlation function although the relative error is still just under 4% and is significantly lower than that for CVM. The better agreement at the exact consolute/critical point is to be expected in view of the fact that $m_4$ has been modified to yield this point correctly.

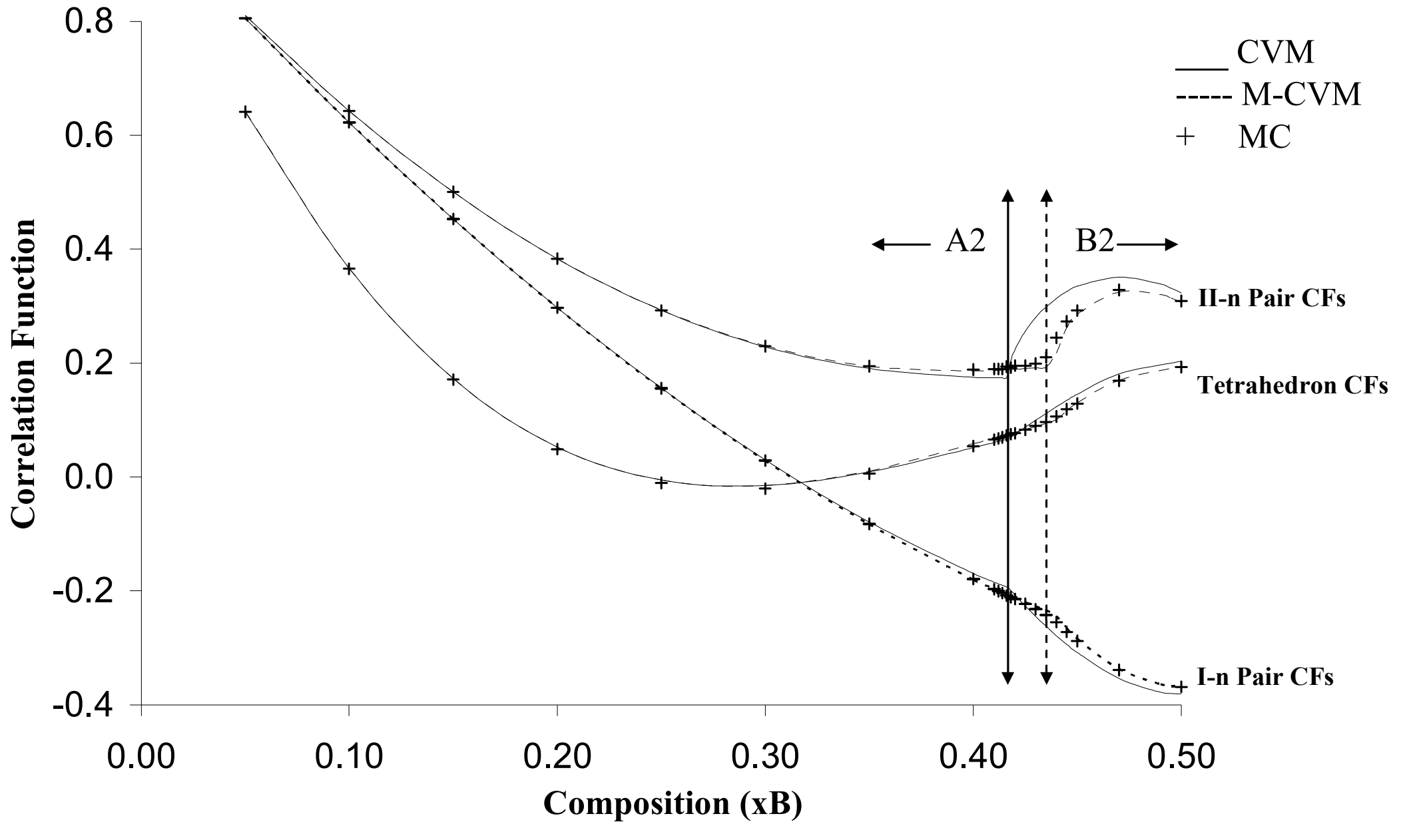

Fig. 4a: Variation of even correlation functions with composition for B2 and A2 phases at $T = 6.2$ kT/J

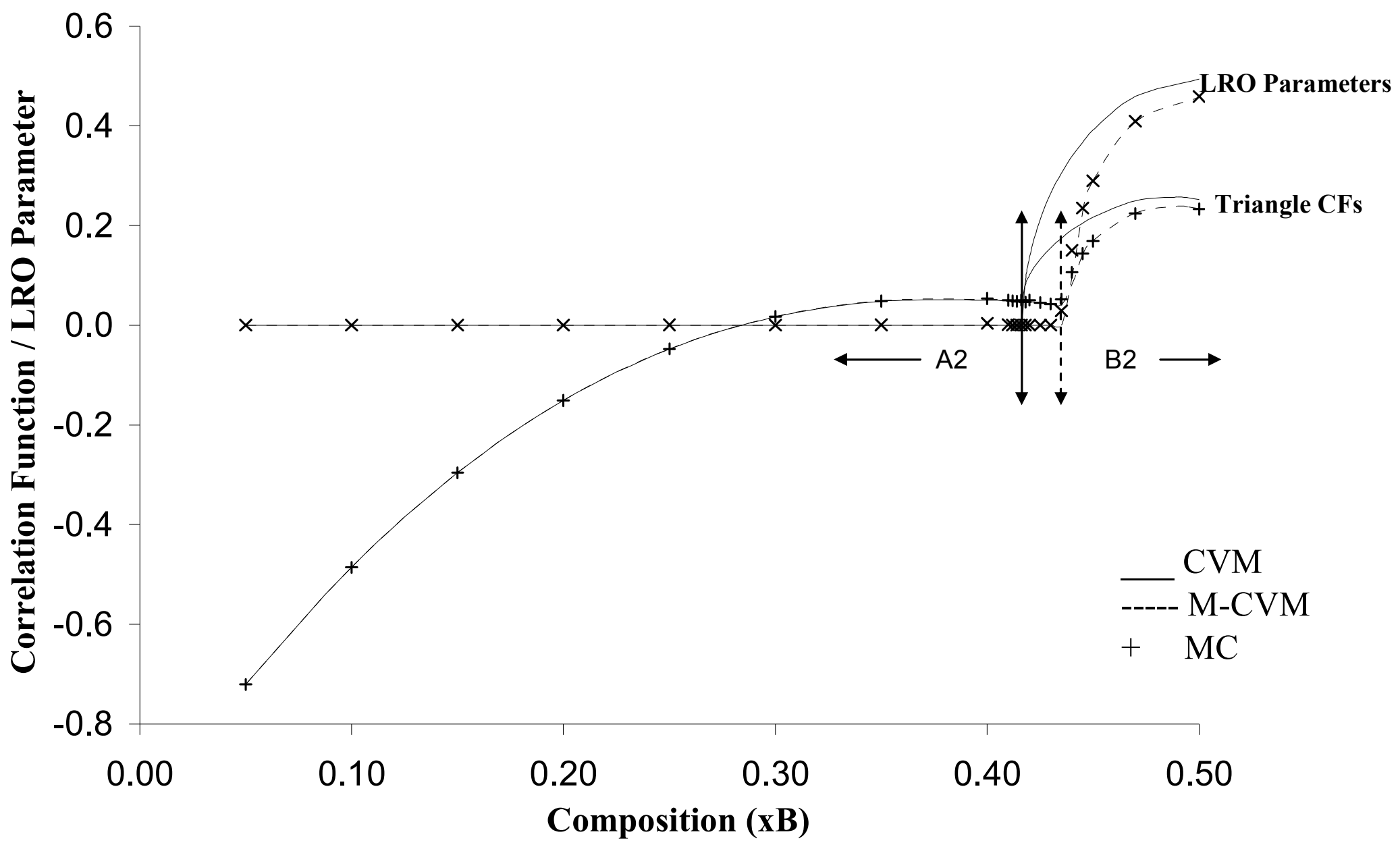

Fig. 4b: Variation of triangle correlation function and long range order parameter with composition for B2 and A2 phases at $T = 6.2$ kT/J

Near the transition composition modification in CVM entropy functional leads to a remarkable improvement in the value of the entropy as shown in fig. 5. For the B2 phase region, calculated values of pair correlation function by modified CVM model are less random than those found by using the CVM functional. On the other hand, the overestimation of pair correlation function values for A2 phase is also corrected with modified model.

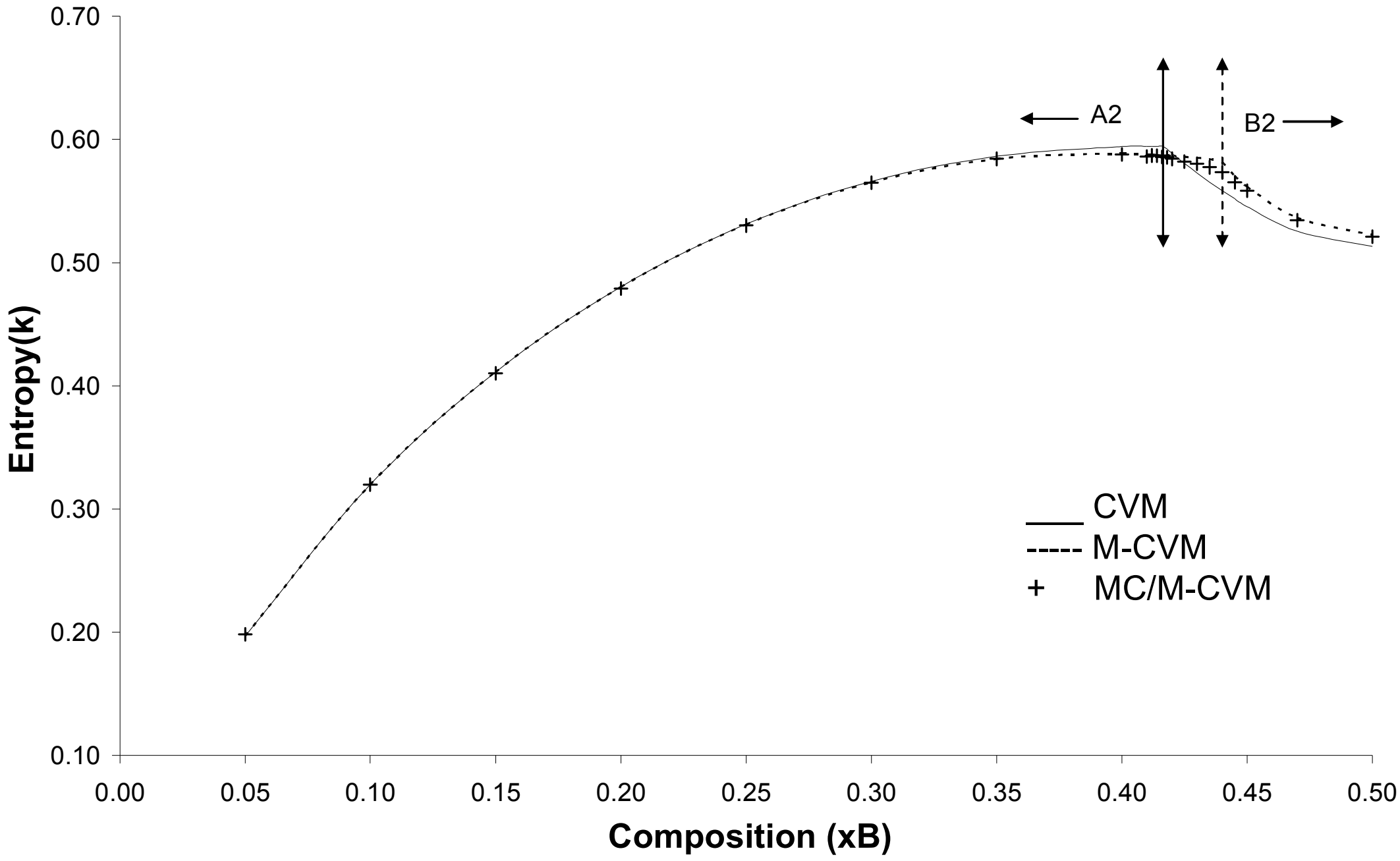

Fig. 5: Variation of entropy with composition for B2 and A2 phases at $T = 6.2$ kT/J

Table 2. Comparison of values of LRO parameter and correlation functions (CF) for $T = 6.2$ kT/J at the A2/B2 phase boundary at $x = 0.432$ for an ordering system and at the spinodal boundary at $x = 0.340$ for a phase separating system

| Structural variable | CVM | M-CVM | MC | Error |
|---|---|---|---|---|
| Ordering system | | | | |
| LRO parameter | 0.2815 | 0.0000 | 0.0059 | 0.0261 |
| I-n pair CF | -0.2516 | -0.2301 | -0.2389 | 0.0010 |
| II-n pair CF ($\alpha\alpha$)<br>II-n pair CF ($\beta\beta$) | 0.2878<br>0.1510 | 0.1924<br>0.1924 | 0.2061<br>0.2033 | 0.0061<br>0.0066 |
| Triangle CF ($\alpha\alpha\beta$)<br>Triangle CF ($\alpha\beta\beta$) | 0.1633<br>-0.0752 | 0.0422<br>0.0422 | 0.0454<br>0.0404 | 0.0112<br>0.0113 |
| Tetrahedron CF | 0.1044 | 0.0909 | 0.0943 | 0.0007 |
| Phase separating system | | | | |
| I-n pair CF | 0.3037 | 0.3184 | 0.3174 | 0.0021 |
| II-n pair CF | 0.2338 | 0.2495 | 0.2500 | 0.0024 |
| Triangle CF | -0.1530 | -0.1599 | -0.1569 | 0.0007 |
| Tetrahedron CF | 0.1450 | 0.1578 | 0.1553 | 0.0017 |

## 4. Discussion

We have shown that the error associated with the truncation of the CVM entropy functional is corrected to a significant extent by a modification of the multiplicity of the term corresponding to the tetrahedron cluster in the entropy functional. The modified multiplicity is found by equating the critical temperatures of ordering as well as phase separating systems to the exactly known. As a consequence the values of correlation functions found by the usual minimization of Gibbs function closely approximate those found from MC simulations even near phase boundaries. We note that this procedure retains the simplicity of CVM and involves no additional computational burden.

There have been several earlier attempts to improve the CVM entropy. The conceptually simplest approach is to use larger basic clusters. Tepesch et al. [7] have used this approach for fcc structures and shown that the values of the correlation functions and thermodynamic functions show greatly improved agreement with those found from MC calculations. The additional computational burden makes this procedure impractical for multi-component systems. The next approach uses a hybrid MC-CVM. Bichara and Inden [6] have used a variant of this method for the bcc structure in which the correlation functions are found by MC simulations and then substituted in the usual CVM entropy functional to find the entropy. As shown by Ferreira et al. [4] in the case of fcc structures, the cancellation of errors in the energy and entropy which occurs in the calculation of the CVM Gibbs function does not occur in this method which partially negates improved agreement. Further, there is an additional computational burden owing to the MC calculations for determining the correlation functions. The other variant of the MC-CVM approach due to Ferreira et al. [4] uses a modified CVM entropy functional applicable to fcc structures. This functional is designed to have the properties of the exact entropy at high and low temperatures. Apart from the additional computational burden for MC calculations of correlation functions, the modification of the entropy functional is solely for the disordered phase and has not kept either the ordered phase or phase separating systems in view. On the other hand, the present approach yields the exact limiting values of entropy from the entropy functionals corresponding to ordered as well as disordered phases, without requiring any MC computations. Further, a computationally efficient approach is the modification of the multiplicity of the basic cluster in the CVM entropy functional. However, in contrast to our approach, Oates et al. [8] have treated this multiplicity as a system dependent parameter in addition to the effective cluster interaction energies for fitting to experimental phase diagram data in an approximate version of CVM called CSA. However, it does not appear appropriate to treat this as a parameter having different values for various systems, since the multiplicity is solely determined by the geometry of the structure.

## Conclusions

(1) An improved CVM entropy functional has been found for the bcc structure by modifying the multiplicity of the basic tetrahedron cluster.

(2) The modified value of multiplicity has been found to be 5.70017 (instead of the usual value of 6) by equating the model transition temperatures with the exact values for ordering as well as phase separating systems.

(3) The correlation functions and the entropy found by using the modified CVM entropy functional are generally within the error bounds of those found from MC calculations.

**Acknowledgements**

The authors would like to thank Dr. Indrajeet Sinha, Department of Applied Chemistry, Institute of Technology, for help in developing the Monte-Carlo code as well as useful discussions.